\documentclass[prb,twocolumn,showpacs]{revtex4}
\usepackage{graphicx} 

\usepackage{amsmath,amsthm,amssymb,mathrsfs}
\usepackage{bm}

\begin{document}

\title{Boltzmann theory of magnetoresistance due to a spin spiral}

\author{Tomohiro Taniguchi$^{1,2}$ and Hiroshi Imamura$^{1}$}
 \affiliation{
 ${}^{1}$ 
 Nanotechnology Research Institute, National Institute of Advanced Industrial Science and Technology, 
 Tsukuba, Ibaraki 305-8568, Japan, \\ 
 ${}^{2}$ 
 Institute of Applied Physics, University of Tsukuba, Tsukuba, Ibaraki 305-8573, Japan
 }

 \date{\today} 
 \begin{abstract}
  {
  We studied the magnetoresistance due to a spin spiral 
  by solving the Boltzmann equation. 
  The scattering rates of conduction electrons are calculated 
  by using the non-perturbative wave function of the conduction electrons
  and the non-equilibrium distribution function is obtained by numerically solving the Boltzmann equation. 
  These enable us to calculate the resistivity of a sufficiently thin spin spiral. 
  A magnetoresistance ratio of more than 50 \% 
  is predicted for a spin spiral with high spin polarization $(\ge\!0.8)$ and 
  a small period (about 1-2 nm). 
  }
 \end{abstract}

 \maketitle



 There is great interest currently in spin-dependent transport phenomena 
 in magnetic domain walls 
 such as the magnetoresistance (MR) effect \cite{gregg96,viret96,ebels00,levy97,simanek01} and 
 spin-transfer torque-driven magnetization dynamics \cite{hayashi06,beach06,zhang04,taniguchi09} 
 because of the potential application of
 these phenomena to spin-electronics devices 
 such as spin-motive-force memory \cite{barnes06} and racetrack memory \cite{parkin08}. 
 In these devices, 
 higher magnetoresistance due to a thin domain wall is desirable 
 for high-density magnetic recording. 


 In 1997, Levy and Zhang \cite{levy97} 
 studied the resistivity due to domain wall scattering 
 by using the same Hamiltonian that was 
 used to explain the giant magnetoresistance effect. 
 They found that the magnetoresistance ratio is proportional to $1/d^{2}$, 
 where $d$ is the thickness of the domain wall, 
 and showed that the magnetoresistance ratio is between 2\% 
 and 11\%, 
 which is consistent with the experimental results (5\%) 
 of Ref. \cite{gregg96} 
 where the thickness of the domain wall is about 15 nm. 
 
 
 However, 
 the theory of Levy and Zhang \cite{levy97} cannot be applied to 
 a sufficiently thin domain wall 
 for two reasons. 
 First, the scattering rates of the conduction electrons are calculated 
 by using the perturbative wave function, 
 which is up to the first order of 
 the dimensionless parameter $\xi$. 
 The parameter $\xi\!=\!l_{J}/d$ characterizes 
 the non-adiabaticity of the spins of the conduction electrons 
 with respect to the localized spins, 
 where $l_{J}\!=\!\pi\hbar v_{\rm F}/(4J)$ is 
 the electrons' traveling length  
 during the precession of their spins around the $sd$-exchange field $J$. 
 For a domain wall with $\xi\!\ge\!1$, 
 the theory cannot estimate the amount of the non-adiabaticity correctly, 
 and thus cannot be applied. 
 Second, since Levy and Zhang applied the diffusion approximation 
 to the Boltzmann equation,
 their theory cannot be applied to the domain wall in the ballistic region 
 $d\!\le\!l_{\rm mfp}$, 
 where $l_{\rm mfp}$ is the mean free path. 
 For conventional ferromagnetic metals such as Fe, Co, Ni, and their alloys, 
 both $l_{J}$ and $l_{\rm mfp}$ are on the order of a few nm \cite{gurney93}.
 

 The thickness of a domain wall is determined by the competition of the
 exchange coupling between the localized magnetizations and the magnetic anisotropy, 
 and is usually on the order of 50 nm for conventional ferromagnetic metals.  
 Recently, however, the production of the domain wall of Co${}_{50}$Fe${}_{50}$, 
 with a thickness of about 2.5 nm, was achieved 
 by trapping the domain wall in a current-confined-path (CCP) geometry \cite{fuke07}, 
 and a magnetoresistance ratio of about 7\%
 -10\% 
 was observed. 
 Many studies have examined 
 to understand the physical properties of the CCP structure 
 and applied that structure to magnetic devices \cite{sato08,matsushita08}. 
 To investigate the transport properties of such a thin magnetic structure, 
 in which the system size $d$ is comparable to or less than $l_{J}$ and $l_{\rm mfp}$, 
 i.e., a few nm, 
 it is important to develop the theory of Levy and Zhang 
 to take into account the amount of the non-adiabaticity correctly 
 and to describe the transport without the diffusion approximation. 
 


 In this paper, 
 we study the dependence of 
 the magnetoresistance ratio of a spin spiral 
 on its period (thickness) $d$ 
 by solving the Boltzmann equation. 
 We extend the theory of Levy and Zhang \cite{levy97} 
 by using the non-perturbative wave function of the conduction electrons 
 in the calculation of the scattering rates and 
 by solving the Boltzmann equation of the non-equilibrium distribution function numerically. 
 These enable us to investigate the resistivity due to a spin spiral with
 $d\!<\!l_{J},l_{\rm mfp}$. 
 We find that the MR ratio is more than 50\% 
 for a spin spiral with high spin polarization $(\beta\!\ge\!0.8)$ 
 and a small period $(d\simeq 1-2 {\rm nm})$. 
 We also find that in the diffusive region, 
 $d\!\ge\!l_{J},l_{\rm mfp}$, 
 the MR ratio is proportional to $1/d^{2}$, 
 while in the ballistic region, $d\!\le\!l_{J},l_{\rm mfp}$, 
 the MR ratio increases with decreasing $d$ 
 more slowly than it does in the diffusive region.


 We consider electron transport in a one-dimensional spin spiral 
 that lies over $-d/2\!\le\! x \!\le\! d/2$, 
 where $d$ is the period of the $\pi$-rotation 
 of the localized spins. 
 We assume that the spin-dependent transport of the conduction electrons is 
 described by the following Hamiltonian:
 \begin{equation}
   \hat{H}_{0}
   \!=\!
   -\frac{\hbar^{2}}{2m} \bm{\nabla}^{2} 
   \!-\!
   J \hat{\bm{\sigma}}\cdot\hat{\mathbf{S}}(\mathbf{r}),
   \label{eq:H_0}
 \end{equation}
 where $J$ is the $sd$-exchange coupling constant 
 between the conduction ($s$-like) electrons and localized ($d$-like) spin, 
 $\hat{\bm{\sigma}}$ is the vector of the Pauli matrices 
 and $\hat{\mathbf{S}}\!=\!(0,-\sin\theta,\cos\theta)$ is the unit vector 
 along the direction of the localized spin. 
 The angle $\theta$ is given by 
 $\theta(x)\!=\!(\pi/d)(x\!+\!d/2)$. 
 On the other hand, 
 the spin-dependent impurity scattering is described by \cite{levy97}
 \begin{equation}
   \hat{V}
   \!=\!
   \sum_{i}
   \left[
     v \!-\! j \hat{\bm{\sigma}}\cdot\hat{\mathbf{S}}(\mathbf{r})
   \right]
   \delta(\mathbf{r}\!-\!\mathbf{R}_{i}),
   \label{eq:V}
 \end{equation}
 where $\mathbf{R}_{i}$ is the position of the impurity, 
 and $v$ and $j$ are the spin-independent and spin-dependent scattering potentials, respectively. 
 The dependence of the transport properties on the direction of the electrons' spin 
 arises from either the exchange energy $J$ or the spin-dependent scattering potential $j$,
 i.e., 
 the spin dependence of the number of the conduction electrons at Fermi level is due to $J$, 
 and the spin dependence of the scattering rate is due to $j$. 
 

 The resistivity of the spin spiral is calculated by solving the Boltzmann equation 
 of the non-equilibrium distribution function $f^{s}(\mathbf{k})$ 
 given by 
 \begin{equation}
 \begin{split}
   -ev_{x}^{s}E 
   \delta(\varepsilon_{\rm F}\!-\!\varepsilon(\mathbf{k},s))
   =&\!\!
   \int\!\!\frac{{\rm d}^{3}\mathbf{k}^{\prime}}{(2\pi)^{3}}
   W_{\mathbf{k}\mathbf{k}^{\prime}}^{ss}
   [f^{s}(\mathbf{k})\!-\!f^{s}(\mathbf{k}^{\prime})]
 \\
   &+\!\!
   \int\!\!\frac{{\rm d}^{3}\mathbf{k}^{\prime}}{(2\pi)^{3}}
   W_{\mathbf{k}\mathbf{k}^{\prime}}^{s-s}
   [f^{s}(\mathbf{k})\!-\!f^{-s}(\mathbf{k}^{\prime})],
   \label{eq:Boltzmann_1}
 \end{split}
 \end{equation}
 where $W_{\mathbf{k}\mathbf{k}^{\prime}}^{ss^{\prime}}$ is the scattering rate 
 of the conduction electrons 
 from the state $(\mathbf{k},s)$ to the state $(\mathbf{k}^{\prime},s^{\prime})$, 
 $\varepsilon_{\rm F}$ is the Fermi energy 
 and $E$ is the strength of the applied electric field. 
 The index $s,s^{\prime}\!=\!\pm$ denotes the eigenstate of $\hat{H}_{0}$ in spin space, 
 which is given by \cite{calvo78} 
 \begin{equation}
   \Psi_{\pm}(\mathbf{r})
   \!=\!
   {\rm e}^{{\rm i}\mathbf{k}\cdot\mathbf{r}}
   \exp
   \left[
     -{\rm i}\frac{\theta(x)}{2}
     \hat{\sigma}_{x}
   \right]
   \exp
   \left[
     -{\rm i}\frac{\phi(k_{x})}{2}
     \hat{\sigma}_{y}
   \right]\!
   \eta_{\pm}\ .
   \label{eq:eigenstate}
 \end{equation}
 Here the angle $\phi(k_{x})$ and the spinor $\eta_{\pm}$ are
 given by 
 \begin{equation}
   \frac{\phi(k_{x})}{2}
   \!=\!
   {\rm arctan}
   \left[
     \frac{k_{x}\theta^{\prime}}{k_{J}^{2}\!+\!\sqrt{(k_{x}\theta^{\prime})^{2}\!+\!k_{J}^{4}}}
   \right],
   \label{eq:tan_phi}
 \end{equation}
 \begin{align}
   \eta_{+}
   \!=\!
   \begin{pmatrix}
     1 \\ 0
   \end{pmatrix}, 
 &&
   \eta_{-}
   \!=\!
   \begin{pmatrix}
     0 \\ 1
   \end{pmatrix},
 \end{align}
 where $\theta^{\prime}\!=\!{\rm d}\theta/{\rm d}x\!=\!\pi/d$ and 
 $k_{J}\!=\!\sqrt{2mJ}/\hbar$, respectively. 
 The factor $\tan(\phi/2)$ characterizes 
 the non-adiabaticity of the spins of the conduction electrons 
 with respect to the localized spins, 
 and is the most important parameter in our calculations.
 It should be noted that this factor is always less than unity 
 for any period $d$ and momentum $k_{x}$. 
 For a sufficiently large period $d$, 
 $\tan(\phi/2)\!\to\!(k_{x}\theta^{\prime})/(2k_{J}^{2})\!=\!(k_{x}/k_{\rm F})\xi$, 
 and the wave function (\ref{eq:eigenstate}) is reduced to 
 the wave function calculated by Levy and Zhang \cite{levy97}. 
 On the other hand, 
 for a small period $d$ where 
 $\xi\!=\!l_{J}/d$ is comparable to or larger than unity, 
 the wave function (\ref{eq:eigenstate}) does not equal 
 the wave function given in Ref. \cite{levy97}. 
 The eigenvalue of $\hat{H}_{0}$ is given by
 \begin{equation}
   \varepsilon(\mathbf{k},s)
   \!=\!
   \frac{\hbar^{2}}{2m}
   \left[
     k^{2}
     \!+\!
     \left(
       \frac{\theta^{\prime}}{2}
     \right)^{2}
     \!-\!
     s
     \sqrt{(k_{x}\theta^{\prime})^{2}\!+\!k_{J}^{4}}
   \right]. 
   \label{eq:energy}
 \end{equation}
 The velocity $v_{x}^{s}$ is given by 
 $v_{x}^{s}\!=\!\partial\varepsilon(\mathbf{k},s)/\partial p_{x}$. 
 The scattering rates 
 are calculated by using the Fermi golden rule with the Born approximation, 
 \begin{equation}
   W_{\mathbf{k}\mathbf{k}^{\prime}}^{ss^{\prime}}
   \!=\!
   \frac{2\pi}{\hbar}
   |V_{\mathbf{k}\mathbf{k}^{\prime}}^{ss^{\prime}}|^{2}
   \delta(\varepsilon(\mathbf{k},s)\!-\!\varepsilon(\mathbf{k}^{\prime},s^{\prime})),
   \label{eq:scattering_rate}
 \end{equation}
 where the matrix elements of the sattering potential (\ref{eq:V}) are calculated 
 by using the wave function (\ref{eq:eigenstate}) and are given by
 \begin{equation}
   |V_{\mathbf{k}\mathbf{k}^{\prime}}^{ss}|^{2}
   \!=\!
   c_{\rm i}\!
   \left[
     (v\!-\!sj) \cos\frac{\phi}{2} \cos\frac{\phi^{\prime}}{2}
     \!+\!
     (v\!+\!sj) \sin\frac{\phi}{2} \sin\frac{\phi^{\prime}}{2}
   \right]^{2},
 \end{equation}
 \begin{equation}
   |V_{\mathbf{k}\mathbf{k}^{\prime}}^{s-s}|^{2}
   \!=\!
   c_{\rm i}\!
   \left[
     (-sv\!+\!j) \cos\frac{\phi}{2} \sin\frac{\phi^{\prime}}{2}
     \!+\!
     (sv\!+\!j) \sin\frac{\phi}{2} \cos\frac{\phi^{\prime}}{2}
   \right]^{2}, 
 \end{equation}
 respectively, where $c_{\rm i}$ is the impurity concentration. 
 Here, for simplicity, we denote $\phi(k_{x})$ and $\phi(k_{x}^{\prime})$ as 
 $\phi$ and $\phi^{\prime}$, respectively. 
 In the limit of $d\!\to\!\infty$, 
 the conduction electrons change the direction of their spins 
 adiabatically, and thus, $\tan(\phi/2)\!\to\!0$ 
 for any momentum $k_{x}$. 
 In this limit, 
 the spin-flip scattering rate is zero,
 i.e., $V_{\mathbf{k}\mathbf{k}^{\prime}}^{s-s}\!=\!0$, 
 and the spin-conserved scattering rate, 
 $W_{\mathbf{k}\mathbf{k}^{\prime}}^{ss}\!\propto\!|V_{\mathbf{k}\mathbf{k}^{\prime}}^{ss}|^{2}$, 
 is independent of the momentum $k_{x}$. 
 On the other hand, 
 in the limit of $d\!\to\!0$, 
 $\tan(\phi/2)\!\to\!1$ for the large momentum $k_{x}\!\simeq\!k_{\rm F}$, 
 which means that the amount of non-adiabaticity is 
 maximized for the conduction electrons with $v_{x}^{s}\!\simeq\!v_{\rm F}$ 
 because the traveling time through the spin spiral of these electrons, 
 $d/v_{x}$, is much shorter than 
 the period of the precession of the spins of the conduction electrons 
 around the exchange field $J$. 
 In Ref. \cite{levy97}, 
 Levy and Zhang approximate that 
 $\cos(\phi/2)\!\to\!1$ and 
 $\sin(\phi/2)\!\to\!\tan(\phi/2)\!\to\!(k_{x}/k_{\rm F})\xi$. 
 It should be noted that 
 for a thin spin spiral 
 where $\xi\!=\!l_{J}/d$ is comparable to or larger than unity, 
 the estimation of the scattering rate 
 $W_{\mathbf{k}\mathbf{k}^{\prime}}^{ss^{\prime}}$ 
 in our theory for large momentum $k_{x}$ is much smaller than 
 that obtained by Levy and Zhang 
 because the factor $\tan(\phi/2)$ in our calculation is 
 always less than unity 
 while the factor $(k_{x}/k_{\rm F})\xi$ used in Ref. \cite{levy97} 
 is larger than unity. 
 Since the resistivity is high for a high scattering rate, 
 the magnetoresistance obtained in our theory is much lower than 
 that obtained by Levy and Zhang, as shown below. 


 To obtain the non-equilibrium distribution function $f^{s}(\mathbf{k})$ 
 from the Boltzmann equation (\ref{eq:Boltzmann_1}), 
 we assume that 
 $f^{s}(\mathbf{k})\!=\!(\partial f^{s(0)}(\mathbf{k})/\partial\varepsilon)g^{s}(\mathbf{k}) \!\simeq\!
  -\delta(\varepsilon_{\rm F}-\varepsilon(\mathbf{k},s))g^{s}(\mathbf{k})$, 
 where $f^{s(0)}(\mathbf{k})$ is the distribution function in equilibrium. 
 Then, Eq. (\ref{eq:Boltzmann_1}) is reduced to 
 \begin{equation}
 \begin{split}
   -ev_{x}^{s}E
   \!=\!&
   -\!\frac{1}{\tau^{s}(k_{x})} g^{s}(k_{x})
   \!+\!
   \frac{m}{2\pi\hbar^{3}}\!
   \int_{-k_{\rm F}^{s}}^{k_{\rm F}^{s}} \!\!{\rm d}k_{x}^{\prime}
   |V_{\mathbf{k}\mathbf{k}^{\prime}}^{ss}|^{2} g^{s}(k_{x}^{\prime})
 \\
   &+\!
   \frac{m}{2\pi\hbar^{3}}\!
   \int_{-k_{\rm F}^{-s}}^{k_{\rm F}^{-s}} \!\!{\rm d}k_{x}^{\prime}
   |V_{\mathbf{k}\mathbf{k}^{\prime}}^{s-s}|^{2} g^{-s}(k_{x}^{\prime}),
   \label{eq:Boltzmann_2}
 \end{split}
 \end{equation}
 where 
 $k_{\rm F}^{s}$ is given by 
 \begin{equation}
   k_{\rm F}^{s}
   \!=\!
   \sqrt{
     k_{\rm F}^{2} \!+\! \left(\frac{\theta^{\prime}}{2}\right)^{2}
     \!+\!
     s\sqrt{(k_{\rm F}\theta^{\prime})^{2}\!+\!k_{J}^{4}}
   }\ .
 \end{equation}
 The relaxation time $\tau^{s}(k_{x})$ is given by 
 $1/\tau^{s}(k_{x})\!=\!1/\tau^{ss}(k_{x})\!+\!1/\tau^{s-s}(k_{x})$, where 
 the spin-conserved relaxation time $\tau^{ss}(k_{x})$ and 
 the spin-flip relaxation time $\tau^{s-s}(k_{x})$ are given by 
 \begin{equation}
   \frac{1}{\tau^{ss^{\prime}}(k_{x})}
   \!=\!
   \frac{m}{2\pi\hbar^{3}}\!
   \int_{-k_{\rm F}^{s^{\prime}}}^{k_{\rm F}^{s^{\prime}}} \!\!{\rm d}k_{x}^{\prime}
   |V_{\mathbf{k}\mathbf{k}^{\prime}}^{ss^{\prime}}|^{2}. 
 \end{equation}
 The distribution function $f^{s}(\mathbf{k})$ is obtained 
 by numerically solving Eq. (\ref{eq:Boltzmann_2}) \cite{penn99}. 
 The resistivity of the spin spiral is calculated as 
 $\rho\!=\!1/(\sigma^{+}\!+\!\sigma^{-})$, 
 where $\sigma^{s}\!=\!-(e/E)\int{\rm d}^{3}\mathbf{k}/(2\pi)^{3} v_{x}^{s}f^{s}(\mathbf{k})$ 
 is the conductivity of the spin-$s$ electrons. 

\begin{figure}
  \centerline{\includegraphics[width=0.8\columnwidth]{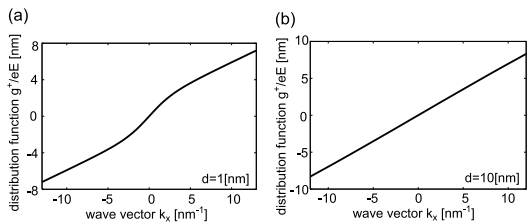}}
  \caption{The dependence of the distribution function, $g^{+}/eE$, 
    on the momentum $k_{x}$ for
    (a) $d\!=\!1$nm and (b) $d\!=\!10$nm, respectively. 
  }
  \label{fig:fig1}
\end{figure}

 In the calculation of the scattering-in term, 
 $\int{\rm d}^{3}\mathbf{k}^{\prime}/(2\pi)^{3}
   [W_{\mathbf{k}\mathbf{k}^{\prime}}^{ss}f^{s}(\mathbf{k}^{\prime})\!
   +\!W_{\mathbf{k}\mathbf{k}^{\prime}}^{s-s}f^{-s}(\mathbf{k}^{\prime})]$,
 in Eq. (\ref{eq:Boltzmann_2}),
 Levy and Zhang \cite{levy97} assume that 
 the non-equilibrium distribution function is proportional to the momentum $k_{x}$. 
 However, we do not apply this diffusion approximation to the scattering-in term 
 because we are interested in the resistivity for a spin spiral with $d\!<\!l_{\rm mfp}$. 
 Figure \ref{fig:fig1} (a) and (b) 
 show typical dependences 
 of the distribution function
 obtained by Eq. (\ref{eq:Boltzmann_2}), $g^{+}/eE$, 
 on the momentum $k_{x}$ 
 for $d\!=\!1$ nm and $d\!=\!10$ nm, respectively, 
 where the mean free path $l_{\rm mfp}$ is taken to be 5.9 nm. 
 According to Fig. \ref{fig:fig1}, 
 we can verify that the diffusion approximation is not applicable 
 to the region $d\!<\!l_{\rm mfp}$ 
 while it is a good approximation to the region $d\!>\!l_{\rm mfp}$. 


 Before estimating the resistivity of a spin spiral, 
 we should emphasize the validity of our calculation. 
 The semi-classical Boltzmann equation is applicable 
 when the system is larger than the width of the wave packet of 
 the conduction electrons, i.e., the Fermi wavelength $\lambda_{\rm F}$. 
 In our calculation, 
 this condition equals $d\!>\!\lambda_{\rm F}$. 
 For conventional ferromagnetic metals, 
 the Fermi wavelength is on the order of a few angstrom, 
 which is one order of magnitude smaller than $l_{J}$ and $l_{\rm mfp}$ \cite{gurney93}. 
 It should also be noted that 
 the derivative of the angle $\theta(x)$ is assumed to be constant 
 in the derivation of the wave function (\ref{eq:eigenstate}). 
 Thus, our calculation is valid for a spin spiral 
 where the direction of the localized spin changes linearly in space. 


  
\begin{figure}
  \centerline{\includegraphics[width=0.65\columnwidth]{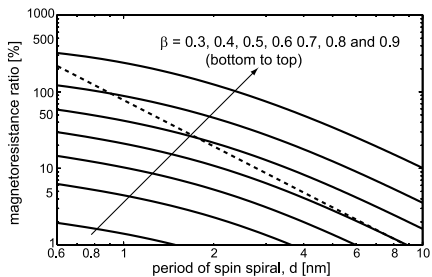}}
  \caption{The dependence of the magnetoresistance (MR) ratio of a spin spiral 
  on its period $d$. 
  The solid lines 
  from bottom to top correspond to the MR ratio with 
  the spin polarizations $\beta\!=$0.3, 0.4, 0.5, 0.6, 0.7, 0.8 and 0.9, respectively. 
  The dashed line 
  is the MR ratio estimated by the theory of Levy and Zhang \cite{levy97} 
  with $\beta\!=\!0.5$. 
  }
  \label{fig:fig2}
\end{figure}

  Figure \ref{fig:fig2} shows the dependence of the MR ratio due to a spin spiral, 
  defined by $(\rho\!-\!\rho^{(0)})/\rho^{(0)}$, on its period $d$. 
  The values of the parameters we use are as follows. 
  The Fermi energy $\varepsilon_{\rm F}$ and the $sd$-exchange coupling constant $J$ are taken to be 
  5.0 eV and 0.5 eV, respectively. 
  The Fermi wavelength $\lambda_{\rm F}$ is estimated to be 5.4 \AA.
  The strengths of the impurity scattering, $v$ and $j$, 
  and the impurity concentration, $c_{\rm i}$, are 
  estimated by the resistivity $\rho^{(0)}$ and 
  the spin polarization $\beta$ of a bulk ferromagnetic metal. 
  The value of $\rho^{(0)}$ is taken to be 150 $\Omega$nm,
  which is a typical value of the conventional ferromagnetic metals \cite{bass07}, 
  while the value of $\beta$ is taken to be from 0.3 to 0.9.
  Using these parameters, $l_{J}\!=\!\pi\hbar v_{\rm F}/(4J)$ is estimated to be 1.4 nm, 
  and the mean free path $l_{\rm mfp}\!=\!(l_{\rm mfp}^{+}\!+\!l_{\rm mfp}^{-})/2$, 
  where
  $l_{\rm mfp}^{s}\!=\!v_{\rm F}^{s}\tau^{s(0)}$, $v_{\rm F}^{s}\!=\!\hbar k_{\rm F}^{s(0)}/m$, 
  $\tau^{s(0)}\!=\!\pi\hbar^{3}/[mc_{\rm i}(v\!-\!sj)^{2}k_{\rm F}^{s(0)}]$,
  and $k_{\rm F}^{s(0)}\!=\!\sqrt{k_{\rm F}^{2}\!+\!sk_{J}^{2}}$, 
  is estimated to be 5.9 nm,
  which is approximately independent of the values of $\beta$. 


  As shown in Fig. \ref{fig:fig2}, 
  the MR ratio increases as the period $d$ decreases. 
  The higher the spin polarization of the bulk $\beta$ is, 
  the higher the MR ratio is. 
  In the diffusive region $d\!>\!l_{J},l_{\rm mfp}$, 
  the MR ratio is estimated to be 1\%-20\%. 
  On the other hand, 
  for a thin spin spiral ($d\!\simeq\!1\!-\!2$nm) with a high polarization 
  $(\beta\!\simeq\!0.8\!-\!0.9)$, 
  an MR ratio of more than 50\% 
  is predicted. 
  Recently, a spin spiral of ferromagnetic Mn/W(001) with the rotation period $2d\!\simeq\! 2.2$ nm 
  was created experimentally \cite{ferriani08}, 
  whose period $d$ is comparable to or smaller than $l_{J}$ and $l_{\rm mfp}$. 
  Thus, it is reasonable to consider such a sufficiently thin spin spiral
  $d\!<\!l_{J},l_{\rm mfp}$. 
  The values of the spin polarization $\beta$ 
  of the conventional ferromagnetic metals 
  such as Fe, Co, Ni, and their alloys are about 0.5-0.7; 
  for example, $\beta\!=\!0.51$ for Co, 
  0.65 for Co${}_{91}$Fe${}_{9}$, 
  and 0.73 for Ni${}_{80}$Fe${}_{20}$ \cite{reilly99,fert99}. 
  The value of $\beta$ depends on 
  the combination and the composition ratio of the ferromagnetic metals, 
  and we can expect ferromagnetic metals 
  with high spin polarizations. 
  Thus, the prediction of our calculation  
  for a spin spiral with high spin polarization $\beta$ 
  and a small period $d\!<\!l_{J},l_{\rm mfp}$ 
  will be confirmed experimentally. 



  The physics behind these results are as follows. 
  The origin of MR due to a spin spiral is 
  the mixing of the channels of the spin-up current and 
  spin-down current due to the spin-dependent 
  scattering potential $\hat{V}$. 
  The channel mixing increases 
  the scattering probability of the conduction electrons,
  and thus the resistivity. 
  The mixing due to the scattering arises from 
  the non-adiabaticity of the spins of the conduction electrons, 
  which is characterized by $\tan[\phi(k_{x})/2]$. 
  In the limit of $d\!\to\!\infty$, 
  the conduction electrons change the direction of their spins adiabatically, 
  i.e., $\tan(\phi/2)\!\to\!0$ for any momentum $k_{x}$, 
  and the MR ratio tends to be zero. 
  On the other hand, 
  in the limit of $d\!\to\!0$,  
  the amount of non-adiabaticity that is maximized 
  for the conduction electrons with large momentum $k_{x}$, 
  i.e., $\tan(\phi/2)\!\to\!1$ for $k_{x}\!\simeq\!k_{\rm F}$, 
  and thus the MR ratio, increase 
  as the period $d$ decreases. 
  In other words, 
  the MR due to the spin spiral is 
  mainly due to the conduction electrons 
  with large momentum $k_{x}$. 
  Since the MR arises from the asymmetry of the transport properties 
  of the spin channels, 
  the higher the spin polarization $\beta$ is, 
  the higher the MR ratio is.


 The dashed line in Fig. \ref{fig:fig2} shows 
 the MR ratio estimated by 
 the theory of Levy and Zhang 
 with $\beta\!=\!0.5$ \cite{comment}; 
 \begin{equation}
   {\rm MR\ ratio}
   \!=\!
   \frac{4}{5}\xi^{2}
   \left(
     \frac{\beta^{2}}{1\!-\!\beta^{2}}
   \right)
   \left(
     3 \!-\! \frac{5\sqrt{1\!-\!\beta^{2}}}{3}
   \right).
 \end{equation}
 By comparing the solid line and the dashed line in Fig. \ref{fig:fig2}, 
 we find that the MR ratio in the diffusive region, 
 $d\!>\!l_{J},l_{\rm mfp}$, is proportional to $1/d^{2}$, 
 as shown by Levy and Zhang \cite{levy97}. 
 On the other hand, 
 in the ballistic region, $d\!<\!l_{J},l_{\rm mfp}$, 
 the MR ratio increases more slowly as the period $d$ decreases
 compared to the diffusive region. 
 It should be noted that 
 the factor $\tan(\phi/2)$ is approximated to be $(k_{x}/k_{\rm F})\xi$ 
 in Ref. \cite{levy97}, 
 which is on the first order of $1/d$. 
 However, for a thin spin spiral, 
 the higher-order terms of $1/d$ also contribute to 
 the calculations of resistivity, 
 and the dependence of the MR ratio on the period $d$ 
 shifts from $1/d^{2}$. 
 As shown in Fig. \ref{fig:fig2}, 
 the MR ratio obtained by our theory is much smaller than 
 that obtained by Levy and Zhang. 
 This is due to the fact that 
 the estimated scattering rate by our calculation is 
 much lower than that by Levy and Zhang, 
 as mentioned above. 
 The smaller the period $d$ is, 
 the larger 
 the difference is in the amount of non-adiabaticity 
 between our theory and that of Levy and Zhang, 
 i.e., the difference in the values of $\tan(\phi/2)$ and $(k_{x}/k_{\rm F})\xi$. 
 Thus, the difference in the MR ratio between 
 our theory and theirs increases 
 as the period $d$ decreases.




 In conclusion, 
 we have studied the dependence of 
 magnetoresistance due to a spin spiral 
 on its period $d$ 
 by solving the Boltzmann equation. 
 The scattering rate of the conduction electrons in the spin spiral is calculated 
 by using the non-perturbative wave function of the conduction electrons,
 and the non-equilibrium distribution function is obtained 
 by numerically solving the Boltzmann equation. 
 An MR ratio of more than 50\% 
 is predicted for a thin spin spiral $(d\!\simeq\!1-2 {\rm nm})$ 
 with high spin polarization $(\beta\!\ge\!0.8)$. 
 We also find that 
 the MR ratio in the diffusive region is proportional to $1/d^{2}$, 
 while in the ballistic region 
 the MR ratio increases more slowly with decreasing $d$ 
 compared to the diffusive region. 
 

 The author would like to acknowledge the valuable discussions they had
 with P. M. Levy, Y. Utsumi, Y. Rikitake, 
 J. Sato, K. Matsushita, N. Yokoshi, 
 and S. Kawasaki. 
 This work was supported by JSPS.





\begin{thebibliography}{22}
\expandafter\ifx\csname natexlab\endcsname\relax\def\natexlab#1{#1}\fi
\expandafter\ifx\csname bibnamefont\endcsname\relax
  \def\bibnamefont#1{#1}\fi
\expandafter\ifx\csname bibfnamefont\endcsname\relax
  \def\bibfnamefont#1{#1}\fi
\expandafter\ifx\csname citenamefont\endcsname\relax
  \def\citenamefont#1{#1}\fi
\expandafter\ifx\csname url\endcsname\relax
  \def\url#1{\texttt{#1}}\fi
\expandafter\ifx\csname urlprefix\endcsname\relax\def\urlprefix{URL }\fi
\providecommand{\bibinfo}[2]{#2}
\providecommand{\eprint}[2][]{\url{#2}}

\bibitem[{\citenamefont{Gregg et~al.}(1996)\citenamefont{Gregg, Allen,
  Ounadjela, Viret, Hehn, Thompson, and Coey}}]{gregg96}
\bibinfo{author}{\bibfnamefont{J.~F.} \bibnamefont{Gregg}},
  \bibinfo{author}{\bibfnamefont{W.}~\bibnamefont{Allen}},
  \bibinfo{author}{\bibfnamefont{K.}~\bibnamefont{Ounadjela}},
  \bibinfo{author}{\bibfnamefont{M.}~\bibnamefont{Viret}},
  \bibinfo{author}{\bibfnamefont{M.}~\bibnamefont{Hehn}},
  \bibinfo{author}{\bibfnamefont{M.}~\bibnamefont{Thompson}}, \bibnamefont{and}
  \bibinfo{author}{\bibfnamefont{J.~M.~D.} \bibnamefont{Coey}},
  \bibinfo{journal}{Phys. Rev. Lett.} \textbf{\bibinfo{volume}{77}},
  \bibinfo{pages}{1580} (\bibinfo{year}{1996}).

\bibitem[{\citenamefont{Viret et~al.}(1996)\citenamefont{Viret, Vignoles, Cole,
  and Coey}}]{viret96}
\bibinfo{author}{\bibfnamefont{M.}~\bibnamefont{Viret}},
  \bibinfo{author}{\bibfnamefont{D.}~\bibnamefont{Vignoles}},
  \bibinfo{author}{\bibfnamefont{D.}~\bibnamefont{Cole}}, \bibnamefont{and}
  \bibinfo{author}{\bibfnamefont{J.~M.~D.} \bibnamefont{Coey}},
  \bibinfo{journal}{Phys. Rev. B} \textbf{\bibinfo{volume}{53}},
  \bibinfo{pages}{8464} (\bibinfo{year}{1996}).

\bibitem[{\citenamefont{Ebels et~al.}(2000)\citenamefont{Ebels, Radulescu,
  Henry, Piraux, and Ounadjela}}]{ebels00}
\bibinfo{author}{\bibfnamefont{U.}~\bibnamefont{Ebels}},
  \bibinfo{author}{\bibfnamefont{A.}~\bibnamefont{Radulescu}},
  \bibinfo{author}{\bibfnamefont{Y.}~\bibnamefont{Henry}},
  \bibinfo{author}{\bibfnamefont{L.}~\bibnamefont{Piraux}}, \bibnamefont{and}
  \bibinfo{author}{\bibfnamefont{K.}~\bibnamefont{Ounadjela}},
  \bibinfo{journal}{Phys. Rev. Lett.} \textbf{\bibinfo{volume}{84}},
  \bibinfo{pages}{983} (\bibinfo{year}{2000}).

\bibitem[{\citenamefont{Levy and Zhang}(1997)}]{levy97}
\bibinfo{author}{\bibfnamefont{P.~M.} \bibnamefont{Levy}} \bibnamefont{and}
  \bibinfo{author}{\bibfnamefont{S.}~\bibnamefont{Zhang}},
  \bibinfo{journal}{Phys. Rev. Lett.} \textbf{\bibinfo{volume}{79}},
  \bibinfo{pages}{5110} (\bibinfo{year}{1997}).

\bibitem[{\citenamefont{Simanek}(2001)}]{simanek01}
\bibinfo{author}{\bibfnamefont{E.}~\bibnamefont{Simanek}},
  \bibinfo{journal}{Phys. Rev. B} \textbf{\bibinfo{volume}{63}},
  \bibinfo{pages}{224412} (\bibinfo{year}{2001}).

\bibitem[{\citenamefont{Hayashi et~al.}(2006)\citenamefont{Hayashi, Thomas,
  Bazaliy, Moriya, Jiang, and Parkin}}]{hayashi06}
\bibinfo{author}{\bibfnamefont{M.}~\bibnamefont{Hayashi}},
  \bibinfo{author}{\bibfnamefont{L.}~\bibnamefont{Thomas}},
  \bibinfo{author}{\bibfnamefont{Y.~B.} \bibnamefont{Bazaliy}},
  \bibinfo{author}{\bibfnamefont{R.}~\bibnamefont{Moriya}},
  \bibinfo{author}{\bibfnamefont{X.}~\bibnamefont{Jiang}}, \bibnamefont{and}
  \bibinfo{author}{\bibfnamefont{S.~S.~P.} \bibnamefont{Parkin}},
  \bibinfo{journal}{Phys. Rev. Lett.} \textbf{\bibinfo{volume}{96}},
  \bibinfo{pages}{197207} (\bibinfo{year}{2006}).

\bibitem[{\citenamefont{Beach et~al.}(2006)\citenamefont{Beach, Knutson,
  Nistor, Tsoi, and Erskin}}]{beach06}
\bibinfo{author}{\bibfnamefont{G.~S.~D.} \bibnamefont{Beach}},
  \bibinfo{author}{\bibfnamefont{C.}~\bibnamefont{Knutson}},
  \bibinfo{author}{\bibfnamefont{C.}~\bibnamefont{Nistor}},
  \bibinfo{author}{\bibfnamefont{M.}~\bibnamefont{Tsoi}}, \bibnamefont{and}
  \bibinfo{author}{\bibfnamefont{L.}~\bibnamefont{Erskin}},
  \bibinfo{journal}{Phys. Rev. Lett.} \textbf{\bibinfo{volume}{97}},
  \bibinfo{pages}{057203} (\bibinfo{year}{2006}).

\bibitem[{\citenamefont{Zhang and Li}(2004)}]{zhang04}
\bibinfo{author}{\bibfnamefont{S.}~\bibnamefont{Zhang}} \bibnamefont{and}
  \bibinfo{author}{\bibfnamefont{Z.}~\bibnamefont{Li}}, \bibinfo{journal}{Phys.
  Rev. Lett.} \textbf{\bibinfo{volume}{93}}, \bibinfo{pages}{127204}
  (\bibinfo{year}{2004}).

\bibitem[{\citenamefont{Taniguchi et~al.}(2009)\citenamefont{Taniguchi, Sato,
  and Imamura}}]{taniguchi09}
\bibinfo{author}{\bibfnamefont{T.}~\bibnamefont{Taniguchi}},
  \bibinfo{author}{\bibfnamefont{J.}~\bibnamefont{Sato}}, \bibnamefont{and}
  \bibinfo{author}{\bibfnamefont{H.}~\bibnamefont{Imamura}},
  \bibinfo{journal}{Phys. Rev. B} \textbf{\bibinfo{volume}{79}},
  \bibinfo{pages}{212410} (\bibinfo{year}{2009}).

\bibitem[{\citenamefont{Barnes et~al.}(2006)\citenamefont{Barnes, Ieda, and
  Maekawa}}]{barnes06}
\bibinfo{author}{\bibfnamefont{S.~E.} \bibnamefont{Barnes}},
  \bibinfo{author}{\bibfnamefont{J.}~\bibnamefont{Ieda}}, \bibnamefont{and}
  \bibinfo{author}{\bibfnamefont{S.}~\bibnamefont{Maekawa}},
  \bibinfo{journal}{Appl. Phys. Lett.} \textbf{\bibinfo{volume}{89}},
  \bibinfo{pages}{122507} (\bibinfo{year}{2006}).

\bibitem[{\citenamefont{Parkin et~al.}(2008)\citenamefont{Parkin, Hayashi, and
  Thomas}}]{parkin08}
\bibinfo{author}{\bibfnamefont{S.~S.~P.} \bibnamefont{Parkin}},
  \bibinfo{author}{\bibfnamefont{M.}~\bibnamefont{Hayashi}}, \bibnamefont{and}
  \bibinfo{author}{\bibfnamefont{L.}~\bibnamefont{Thomas}},
  \bibinfo{journal}{Science} \textbf{\bibinfo{volume}{320}},
  \bibinfo{pages}{190} (\bibinfo{year}{2008}).

\bibitem[{\citenamefont{Gurney et~al.}(1993)\citenamefont{Gurney, Speriosu,
  Nozieres, Lefakis, Wilhoit, and Need}}]{gurney93}
\bibinfo{author}{\bibfnamefont{B.~A.} \bibnamefont{Gurney}},
  \bibinfo{author}{\bibfnamefont{V.~S.} \bibnamefont{Speriosu}},
  \bibinfo{author}{\bibfnamefont{J.-P.} \bibnamefont{Nozieres}},
  \bibinfo{author}{\bibfnamefont{H.}~\bibnamefont{Lefakis}},
  \bibinfo{author}{\bibfnamefont{D.~R.} \bibnamefont{Wilhoit}},
  \bibnamefont{and} \bibinfo{author}{\bibfnamefont{O.~U.} \bibnamefont{Need}},
  \bibinfo{journal}{Phys. Rev. Lett.} \textbf{\bibinfo{volume}{71}},
  \bibinfo{pages}{4023} (\bibinfo{year}{1993}).

\bibitem[{\citenamefont{Fuke et~al.}(2007)\citenamefont{Fuke, Hashimoto,
  Takagishi, Iwasaki, Kawasaki, Miyake, and Sahashi}}]{fuke07}
\bibinfo{author}{\bibfnamefont{H.~N.} \bibnamefont{Fuke}},
  \bibinfo{author}{\bibfnamefont{S.}~\bibnamefont{Hashimoto}},
  \bibinfo{author}{\bibfnamefont{M.}~\bibnamefont{Takagishi}},
  \bibinfo{author}{\bibfnamefont{H.}~\bibnamefont{Iwasaki}},
  \bibinfo{author}{\bibfnamefont{S.}~\bibnamefont{Kawasaki}},
  \bibinfo{author}{\bibfnamefont{K.}~\bibnamefont{Miyake}}, \bibnamefont{and}
  \bibinfo{author}{\bibfnamefont{M.}~\bibnamefont{Sahashi}},
  \bibinfo{journal}{IEEE. Trans. Mag.} \textbf{\bibinfo{volume}{43}},
  \bibinfo{pages}{2848} (\bibinfo{year}{2007}).

\bibitem[{\citenamefont{Sato et~al.}(2008)\citenamefont{Sato, Matsushita, and
  Imamura}}]{sato08}
\bibinfo{author}{\bibfnamefont{J.}~\bibnamefont{Sato}},
  \bibinfo{author}{\bibfnamefont{K.}~\bibnamefont{Matsushita}},
  \bibnamefont{and} \bibinfo{author}{\bibfnamefont{H.}~\bibnamefont{Imamura}},
  \bibinfo{journal}{IEEE. Trans. Mag.} \textbf{\bibinfo{volume}{44}},
  \bibinfo{pages}{2608} (\bibinfo{year}{2008}).

\bibitem[{\citenamefont{Matsushita et~al.}(2008)\citenamefont{Matsushita, Sato,
  and Imamura}}]{matsushita08}
\bibinfo{author}{\bibfnamefont{K.}~\bibnamefont{Matsushita}},
  \bibinfo{author}{\bibfnamefont{J.}~\bibnamefont{Sato}}, \bibnamefont{and}
  \bibinfo{author}{\bibfnamefont{H.}~\bibnamefont{Imamura}},
  \bibinfo{journal}{IEEE. Trans. Mag.} \textbf{\bibinfo{volume}{44}},
  \bibinfo{pages}{2616} (\bibinfo{year}{2008}).

\bibitem[{\citenamefont{Calvo}(1978)}]{calvo78}
\bibinfo{author}{\bibfnamefont{M.}~\bibnamefont{Calvo}},
  \bibinfo{journal}{Phys. Rev. B} \textbf{\bibinfo{volume}{18}},
  \bibinfo{pages}{5073} (\bibinfo{year}{1978}).

\bibitem[{\citenamefont{Penn and Stiles}(1999)}]{penn99}
\bibinfo{author}{\bibfnamefont{D.~R.} \bibnamefont{Penn}} \bibnamefont{and}
  \bibinfo{author}{\bibfnamefont{M.~D.} \bibnamefont{Stiles}},
  \bibinfo{journal}{Phys. Rev. B} \textbf{\bibinfo{volume}{59}},
  \bibinfo{pages}{13338} (\bibinfo{year}{1999}).

\bibitem[{\citenamefont{Bass and W.~P.~Pratt}(2007)}]{bass07}
\bibinfo{author}{\bibfnamefont{J.}~\bibnamefont{Bass}} \bibnamefont{and}
  \bibinfo{author}{\bibfnamefont{J.}~\bibnamefont{W.~P.~Pratt}},
  \bibinfo{journal}{J. Phys.: Condens. Matter} \textbf{\bibinfo{volume}{19}},
  \bibinfo{pages}{183201} (\bibinfo{year}{2007}).

\bibitem[{\citenamefont{Ferriani et~al.}(2008)\citenamefont{Ferriani, von
  Bergmann, Vedmedenko, Heinze, Bode, Heide, Bihlmayer, Blugel, and
  Wiesendanger}}]{ferriani08}
\bibinfo{author}{\bibfnamefont{P.}~\bibnamefont{Ferriani}},
  \bibinfo{author}{\bibfnamefont{K.}~\bibnamefont{von Bergmann}},
  \bibinfo{author}{\bibfnamefont{E.~Y.} \bibnamefont{Vedmedenko}},
  \bibinfo{author}{\bibfnamefont{S.}~\bibnamefont{Heinze}},
  \bibinfo{author}{\bibfnamefont{M.}~\bibnamefont{Bode}},
  \bibinfo{author}{\bibfnamefont{M.}~\bibnamefont{Heide}},
  \bibinfo{author}{\bibfnamefont{G.}~\bibnamefont{Bihlmayer}},
  \bibinfo{author}{\bibfnamefont{S.}~\bibnamefont{Blugel}}, \bibnamefont{and}
  \bibinfo{author}{\bibfnamefont{R.}~\bibnamefont{Wiesendanger}},
  \bibinfo{journal}{Phys. Rev. Lett.} \textbf{\bibinfo{volume}{101}},
  \bibinfo{pages}{027201} (\bibinfo{year}{2008}).

\bibitem[{\citenamefont{Reilly et~al.}(1999)\citenamefont{Reilly, Park, Slater,
  Ouaglal, Lololee, and Jr.}}]{reilly99}
\bibinfo{author}{\bibfnamefont{A.~C.} \bibnamefont{Reilly}},
  \bibinfo{author}{\bibfnamefont{W.}~\bibnamefont{Park}},
  \bibinfo{author}{\bibfnamefont{R.}~\bibnamefont{Slater}},
  \bibinfo{author}{\bibfnamefont{B.}~\bibnamefont{Ouaglal}},
  \bibinfo{author}{\bibfnamefont{R.}~\bibnamefont{Lololee}}, \bibnamefont{and}
  \bibinfo{author}{\bibfnamefont{W.~P.~P.} \bibnamefont{Jr.}},
  \bibinfo{journal}{J. Magn. Magn. Mater.} \textbf{\bibinfo{volume}{195}},
  \bibinfo{pages}{L269} (\bibinfo{year}{1999}).

\bibitem[{\citenamefont{Fert and Piraux}(1999)}]{fert99}
\bibinfo{author}{\bibfnamefont{A.}~\bibnamefont{Fert}} \bibnamefont{and}
  \bibinfo{author}{\bibfnamefont{L.}~\bibnamefont{Piraux}},
  \bibinfo{journal}{J. Magn. Magn. Mater.} \textbf{\bibinfo{volume}{200}},
  \bibinfo{pages}{338} (\bibinfo{year}{1999}).

\bibitem[{com()}]{comment}
\bibinfo{note}{The original paper of Levy and Zhang (Ref. 4) 
  contains a typographic error in the coefficient of the second
  term on the right-hand side. 
  In their paper, the coefficient is 5, not -5/3.}

\end{thebibliography}

\end{document}